\documentstyle[prl,aps,epsfig]{revtex}        

\begin{document}
\catcode`\ä = \active \catcode`\ö = \active \catcode`\ü = \active
\catcode`\Ä = \active \catcode`\Ö = \active \catcode`\Ü = \active
\catcode`\ß = \active \catcode`\é = \active \catcode`\è = \active
\catcode`\ë = \active \catcode`\ô = \active \catcode`\ê = \active
\catcode`\ø = \active \catcode`\ò = \active \catcode`\í = \active
\defä{\"a} \defö{\"o} \defü{\"u} \defÄ{\"A} \defÖ{\"O} \defÜ{\"U} \defß{\ss} \defé{\'{e}}
\defè{\`{e}} \defë{\"{e}} \defô{\^{o}} \defê{\^{e}} \defø{\o} \defò{\`{o}} \defí{\'{i}}
\draft               
\twocolumn[\hsize\textwidth\columnwidth\hsize\csname
@twocolumnfalse\endcsname 

\title{Radio-Frequency Spectroscopy of Ultracold Fermions} \vspace{-5mm}
\author{S. Gupta$^1$, Z. Hadzibabic$^1$, M.W. Zwierlein$^1$, C.A. Stan$^1$, K. Dieckmann$^1$,
C.H. Schunck$^1$, E.G.M. van Kempen$^2$, B.J. Verhaar$^2$ and W.
Ketterle$^1$}
\address{$^1$Department of Physics, MIT-Harvard Center for Ultracold
Atoms, and Research Laboratory of Electronics, MIT, Cambridge, MA
02139}
\address{$^2$ Eindhoven University of Technology,
P.O. Box 513, 5600 MB Eindhoven, The Netherlands}
\maketitle

\begin{abstract}
Radio-frequency techniques were used to study ultracold fermions.
We observed the absence of mean-field ``clock" shifts, the
dominant source of systematic error in current atomic clocks based
on bosonic atoms. This is a direct consequence of fermionic
antisymmetry. Resonance shifts proportional to interaction
strengths were observed in a three-level system. However, in the
strongly interacting regime, these shifts became very small,
reflecting the quantum unitarity limit and many-body effects. This
insight into an interacting Fermi gas is relevant for the quest to
observe superfluidity in this system.
\end{abstract}
 \vskip1pc]

\narrowtext


Radio-frequency (RF) spectroscopy of ultracold atoms provides the
standard of time. However, the resonance frequencies are sensitive
to the interactions between atoms, leading to the so-called clock
shifts of the unperturbed resonances \cite{gibb93}. These shifts
limit the accuracy of current atomic clocks \cite{fert00,sort01},
but can also be used to characterize atomic interactions.

RF spectroscopy has previously been applied to cold atoms to
determine the size and temperature of atom clouds
\cite{mart88,bloc99}. RF methods have also been used for
evaporative cooling, for preparing spinor Bose-Einstein
condensates (BEC) \cite{matt98,sten98}, and as an output coupler
for atom lasers \cite{bloc99,mewe97}. In all these experiments,
shifts and broadenings due to atomic interactions were negligible.
Recently, density-dependent frequency shifts of RF transitions
were observed in rubidium \cite{harb02} and sodium \cite{gorl03}
BECs. These frequency shifts are proportional to the difference in
mean-field energies of two internal atomic states and allow
scattering lengths to be extracted. Mean field shifts in BECs have
been observed also by optical spectroscopy \cite{kill98,sten99}.

Here, we apply RF spectroscopy to ultracold clouds of fermions and
demonstrate several phenomena: (1) the absence of a clock shift in
a two-level system due to fermionic antisymmetry, (2) the
emergence of mean-field shifts in a three-level system after the
relaxation of pair correlations, (3) the limitation of mean-field
shifts due to the unitarity limit, and (4) the universality of the
interaction energy in a dense cloud, independent of the attractive
or repulsive nature of the two-particle interactions.

Research in ultracold fermions has advanced rapidly, with six
groups now having reached quantum degeneracy
\cite{dema99,trus01,schr01,gran02,hadz02,roat02}. A major goal is
to induce strong interactions by tuning magnetic fields to
scattering resonances (called Feshbach resonances). Under these
conditions, Cooper pairs of fermions may form, leading to
superfluidity. This would establish a model system for studying
Bardeen-Cooper-Schrieffer (BCS) pairing at densities nine orders
of magnitude lower than in previous realizations in $^3$He and
superconductors. We show that RF spectroscopy can be used to
characterize interactions between fermions in the regime where
superfluidity has been predicted \cite{houb99,holl01}.



Our experimental technique for preparing ultracold fermions has
been considerably improved since our earlier work
\cite{hadz02,diec02}. As the Pauli exclusion principle suppresses
elastic collisions between identical fermions at low temperatures
and prevents evaporative cooling, we cooled fermionic $^{6}\rm
{Li}$ sympathetically with bosonic $^{23}{\rm Na}$ loaded into the
same magnetic trap. In contrast to previous work, we cooled both
species in their upper hyperfine states ($^{23}{\rm
Na}\!:|F,m_F\rangle = |2,+2\rangle,\,^{6}{\rm Li}\!:|F,m_F\rangle
= |3/2,+3/2\rangle$). This led to a reduction of inelastic loss
processes, boosting our final fermion atom numbers by 2 orders of
magnitude. We could produce BECs containing up to 10 million
sodium atoms in the $|2,+2\rangle$ state by evaporatively cooling
pure bosonic samples in the magnetic trap. For a Bose-Fermi
mixture, the finite heat capacity of the bosons limited the final
lithium temperature after the $30\,$s evaporation cycle to $\sim
\! 0.3\,T_F$ for 10 million fermions and $\sim \! T_F$ for 50
million fermions \cite{hadz03}, where $T_F$ is the Fermi
temperature.

The spin states of $^6$Li of most interest for superfluid pairing
are the two lowest states $|1\rangle$ and $|2\rangle$
($|1/2,+1/2\rangle$ and $|1/2,-1/2\rangle$ at low field), which
are predicted to have an inter-state s-wave Feshbach resonance at
$\sim 800\,$G \cite{houb98,ohar02}. However, both states are
high-field seeking at these fields, making them unsuitable for
magnetic trapping. We therefore transferred the atoms into an
optical trap. For these experiments, 6-8 million
$|3/2,+3/2\rangle$ lithium atoms were loaded into the optical trap
at $T \! \sim \! T_F \sim 35\,\mu$K \cite{ODTload}. The atoms were
then transferred to the lowest energy state $|1\rangle$, using an
adiabatic frequency sweep around the lithium hyperfine splitting
of $228\,$MHz. DC magnetic fields of up to $\sim 900\,$G were
applied, a range encompassing the $|1\rangle-|2\rangle$ Feshbach
resonance. Using RF-induced transitions near $80\,$MHz, we could
create mixtures of states $|1\rangle$, $|2\rangle$ and $|3\rangle$
($|3/2,-3/2\rangle$ at low field), and explore interactions
between fermions in these states.



Collisions between atoms cause a shift of their energy, which is
usually described by the mean-field effect of all the other atoms
on the atom of interest. For example, atoms in state $|2\rangle$
experience an energy shift ${4 \pi \hbar^2 \over m} n_1 a_{12}$
due to the presence of atoms in state $|1\rangle$. Here $\hbar$ is
Planck's constant divided by $2\pi$, $m$ is the mass of the atom,
$n_1$ is the density of $|1\rangle$ atoms and $a_{12}$ is the
inter-state scattering length. We use the convention that positive
scattering length corresponds to a repulsive interaction.
Density-dependent shifts of the resonance frequency for the
transition connecting two states have been observed in
laser-cooled \cite{gibb93} and Bose-condensed clouds
\cite{harb02,gorl03}.

In the case of ultracold fermions, only interactions between
different internal states are allowed. For a system of density
$n$, let us compare the energy of a gas prepared purely in state
$|1\rangle$, to a gas in which one atom is transferred into state
$|2\rangle$. The energy difference is $h \nu_{12} + {4 \pi \hbar^2
\over m} n a_{12}$, where $\nu_{12}$ is the resonance frequency of
the non-interacting system. Similarly, the energy difference
between a gas prepared purely in state $|2\rangle$, and a gas in
which one atom is transferred into state $|1\rangle$ is $h
\nu_{12} - {4 \pi \hbar^2 \over m} n a_{12}$.

However, these energy shifts should not affect the resonance for a
coherent transfer out of a pure state. For fermions in the initial
pure state, the pair-correlation function vanishes at zero
distance due to the antisymmetry of the wavefunction. During any
coherent transfer process, the state vectors of all the atoms
rotate ``in parallel" in Hilbert space, i.e. the superposition of
the two spin states has the same relative phase for all atoms.
Thus, the atoms remain identical and cannot interact in the s-wave
regime. The mean-field energy is thus established only after the
coherence of the superposition state is lost and the pair
correlations have relaxed, forming a purely statistical mixture of
the two states.

It is a consequence of Fermi statistics that spectroscopic methods
do not measure the equilibrium energy difference between the
initial and final state of the system, but rather the unperturbed
resonance frequency. The expected absence of the clock shift has
led to suggestions for the use of fermions in future atomic clocks
\cite{gibb95}. Our work presents an experimental demonstration of
this phenomenon.

\begin{figure}[htbf]
\begin{center}
\vskip0mm \epsfxsize=80mm {\epsfbox{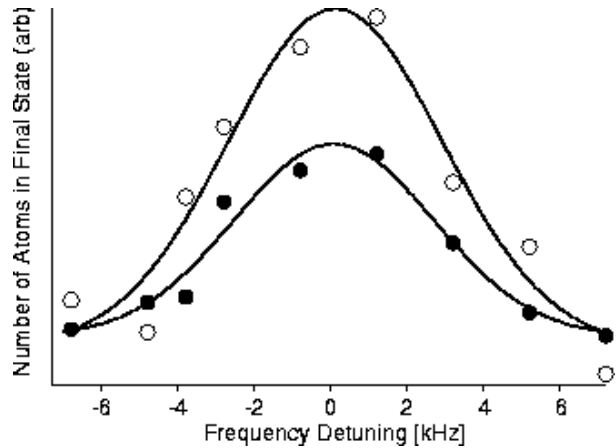}} \vskip1mm
\end{center}
\caption{Absence of the clock shift. RF transitions were driven
between states $|1\rangle$ and $|2\rangle$ on a system prepared
purely in state $|1\rangle$ (filled circles), and purely in state
$|2\rangle$ (open circles). Mean-field interactions would result
in $5\,$kHz shifts for the two curves in opposite directions.
Gaussian fits (solid lines) to the data are separated by $0.04 \pm
0.35\,$kHz. This gives a clock shift suppression factor of 30.}
\label{fig:clock}
\end{figure}

We determined the transition frequency between states $|1\rangle$
and $|2\rangle$, first starting with a pure state $|1\rangle$, and
then with a pure state $|2\rangle$ sample. The absence of a
splitting between these two lines proves the suppression of the
clock shift. Fig.$\,$\ref{fig:clock} shows an example of such
measurements. The magnetic field was ramped up to $570\,$G with
the cloud in state $|1\rangle$. At this field, $a_{12} \sim
150\,a_0$. Therefore, the expected equilibrium mean-field shifts
were $\Delta \nu = \pm 5\,$kHz for our mean density of $3 \times
10^{13}\,$cm$^{-3}$ \cite{density}. The interaction between states
$|1\rangle$ and $|2\rangle$ at this magnetic field was also
observed in the mutual evaporative cooling of the two states in
the optical trap. RF pulses of $140\,\mu$s duration were applied
at frequencies near the unperturbed resonance
$\nu_{12}\sim76\,$MHz. Atoms in states $|1\rangle$ and $|2\rangle$
could be monitored separately by absorption imaging, since they
are optically resolved at this field. We observed a suppression of
the clock shift by a factor of 30 (Fig.$\,$\ref{fig:clock}). Using
the same method, the absence of the clock shift was observed at
several other magnetic fields. In particular, we observed a
suppression of more than three orders of magnitude at $\sim
860\,$G \cite{clockat860}.

P-wave interactions could lead to a non-vanishing clock shift.
However, at these low temperatures they are proportional to $T$ or
$T_F$, whichever is higher, and are therefore strongly suppressed.




Mean-field shifts and scattering lengths can be observed
spectroscopically by driving transitions from a statistical
mixture of two states to a third energy level. (While this work
was in progress, use of a similar method to measure scattering
lengths in fermionic $^{40}$K was reported \cite{rega03}.)
Specifically, we recorded the difference between the RF spectra
for the $|2\rangle \rightarrow |3\rangle$ transition in the
presence and in the absence of state $|1\rangle$ atoms. The
presence of atoms in state $|1\rangle$ is then expected to shift
the resonance by \cite{clock23}:

\begin{equation}\label{meanfield}
  \Delta \nu = {2\hbar \over m}n_1(a_{13}-a_{12})
\end{equation}

\begin{figure}[htbf]
\begin{center}
\vskip0mm \epsfxsize=80mm {\epsfbox{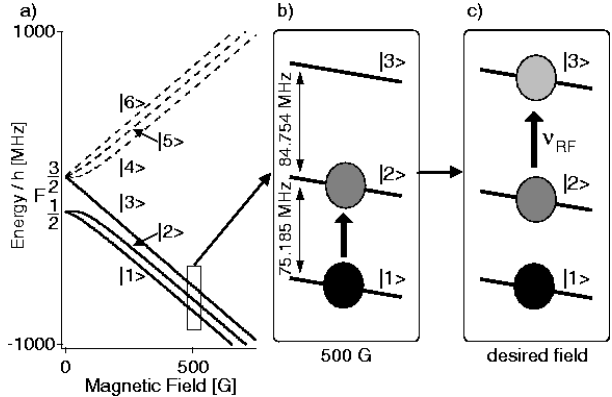}} \vskip1mm
\end{center}
\caption{Schematic of the mean-field measurement. (a) Hyperfine
structure of the ground state of $^6$Li. (b and c) Experimental
scheme: (b) preparation of a mixture of atoms in states
$|1\rangle$ and $|2\rangle$, and (c) RF spectroscopy of the
$|2\rangle \rightarrow |3\rangle$ transition.}
\label{fig:BreitRabi}
\end{figure}

In our experimental scheme to determine the interaction energy at
different magnetic fields (Fig.$\,$\ref{fig:BreitRabi}), the
system was prepared by ramping up the magnetic field to $500\,$G
with the atoms in state $|1\rangle$. Either partial or complete RF
transfer to state $|2\rangle$ was then performed. The number of
atoms in state $|1\rangle$ was controlled by adjusting the speed
of a frequency sweep around the $|1\rangle \! \rightarrow \!
|2\rangle$ resonance. A fast, non-adiabatic sweep created a
superposition of the two states, whereas a slow, adiabatic sweep
prepared the sample purely in state $|2\rangle$. A wait time of
$200\,$ms (see below) was allowed for the coherence between states
$|1\rangle$ and $|2\rangle$ to decay and the system to
equilibrate.

\begin{figure}[htbf]
\begin{center}
\vskip0mm \epsfxsize=80mm {\epsfbox{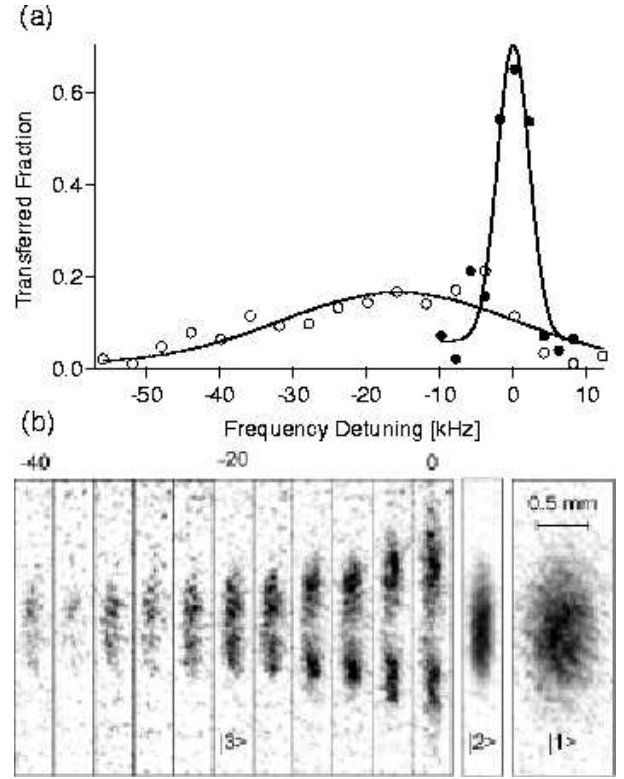}} \vskip1mm
\end{center}
\caption{Representative spectra at $480\,$G. (a) Fraction of atoms
transferred from $|2\rangle$ to $|3 \rangle$, with $|1\rangle$
atoms absent (filled circles), and present (open circles). The
mean-field shift is computed from gaussian fits to the data (solid
lines). (b) Spatial images of state $|3 \rangle$ for the perturbed
resonance. The optical trap was turned off immediately after the
RF pulse and absorption images of the atoms were taken after
$120\,\mu$s expansion time. The central section of $\sim
150\,\mu$m vertical extent was used to extract the transferred
fractions in (a). (b) also shows images of states $|2\rangle$ and
$|1\rangle$ for zero RF detuning. States $|3\rangle$ and
$|2\rangle$ were imaged simultaneously to observe their
complementary spatial structure. State $|1\rangle$ was imaged
after $760\,\mu$s expansion time to record its density for
normalization purposes.} \label{fig:480shift}
\end{figure}

Typical parameters for the decohered $|1\rangle-|2\rangle$ mixture
were mean-density $n_1 \sim 2.4\times 10^{13}\,$cm$^{-3}$ and
$T\!\sim \! 0.7\,T_F$. The magnetic field was then changed to the
desired value, and the transition from state $|2\rangle$ to state
$|3\rangle$ was driven with $140\,\mu$s RF pulses
(Fig.$\,$\ref{fig:BreitRabi}(c)). We monitored the appearance of
atoms in state $|3\rangle$ and the disappearance of atoms from
state $|2\rangle$, using simultaneous absorption imaging.
Fig.$\,$\ref{fig:480shift}(a) shows the unperturbed and perturbed
resonances at $B=480\,$G. The position of the unperturbed
resonance $\nu_{23}$ also determines the magnetic field to
$<0.1\,$G accuracy. Fig.$\,$\ref{fig:480shift}(b) shows absorption
images of atoms in state $|3\rangle$, obtained for different
values of the applied radio-frequency. One can clearly see the
spatial, and thus the density dependence of the mean-field shift:
close to the unperturbed resonance, the low density wings of the
cloud are predominantly transferred, whereas the high density
central part of the cloud is transferred only at sufficient
detuning. To suppress spurious effects from this spatial
dependence, only a small central part of the images was used to
extract the transferred atomic fraction.

\begin{figure}[htbf]
\begin{center}
\vskip0mm \epsfxsize=80mm {\epsfbox{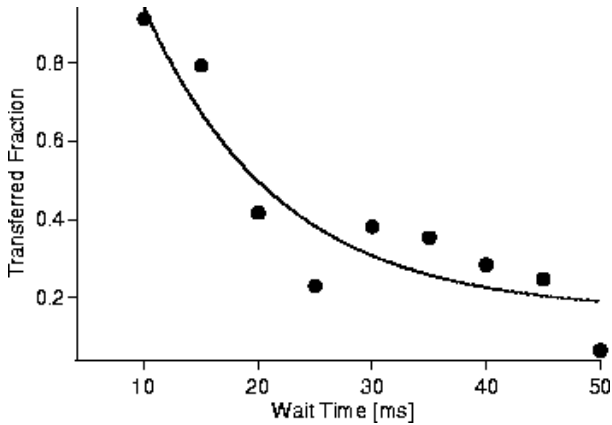}}
\vskip1mm
\end{center}
\caption{Emergence of mean-field shifts due to decoherence at
500$\,$G. Decoherence leads to a reduction of the $|2\rangle
\rightarrow |3\rangle$ transfer at the unperturbed resonance
$\nu_{23}$. An exponential fit to the data (solid line) gives a
time constant of $12\,$ms.} \label{fig:decoherence}
\end{figure}

To ensure that our mean-field measurements were performed on a
statistical mixture, we measured the timescale for decoherence in
our system. The decay of the $|1\rangle-|2\rangle$ coherence at
$500\,$G was observed by monitoring the $|2\rangle \rightarrow
|3\rangle$ transfer at the measured unperturbed resonance
$\nu_{23}$, as a function of wait time
(Fig.$\,$\ref{fig:decoherence}). For wait times small compared to
the decoherence time of the $|1\rangle-|2\rangle$ superposition,
the $|2\rangle \rightarrow |3\rangle$ RF drive places each atom in
an identical three-state superposition. All mean-field shifts are
then absent and the resulting transfer is unchanged from the
unperturbed case. For longer wait times, the $|1\rangle-|2\rangle$
superposition decoheres and mean-field interactions set in. This
shifts the resonance frequency of the $|2\rangle \rightarrow
|3\rangle$ transition, reducing the transferred fraction at
$\nu_{23}$. The measured decoherence time of $\sim 12\,$ms was
attributed mainly to the sensitivity of $\nu_{12}$ to magnetic
field variations across the cloud. These inhomogeneities cause the
relative phase of the $|1\rangle-|2\rangle$ superposition in
different parts of the trap to evolve at different rates, given by
the local $\nu_{12}$. Atoms which travel along different paths
within the trap therefore acquire different phases between their
$|1\rangle$ and $|2\rangle$ components. Being no longer in
identical states, s-wave interactions between them are allowed.
The inhomogeneities scale with the applied magnetic field $B$
while the sensitivity of the transition scales with $\partial
\nu_{12}/\partial B$. We would thus expect the decoherence time to
vary inversely with the product of these two quantities. Our
hypothesis is supported by our observation of longer decoherence
times at higher fields, where $B \times \partial \nu_{12}/\partial
B$ is lower.

\begin{figure}[htbf]
\begin{center}
\vskip0mm \epsfxsize=80mm {\epsfbox{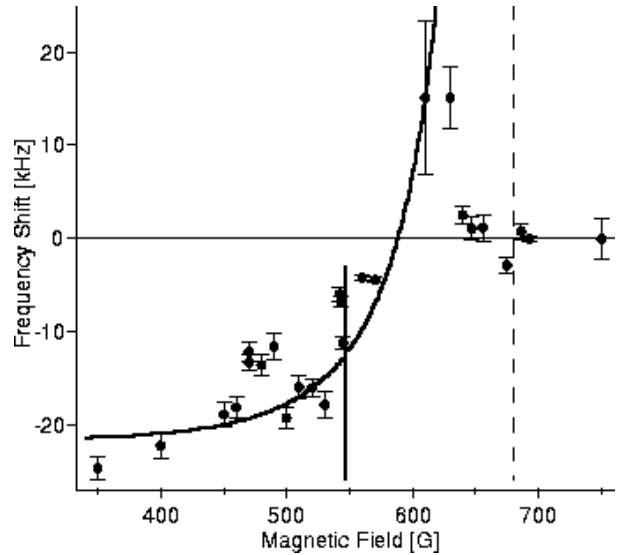}} \vskip1mm
\end{center}
\caption{Frequency shift vs. magnetic field for the $|2\rangle
\rightarrow |3\rangle$ resonance due to atoms in state
$|1\rangle$. The shifts are computed by monitoring the arrival
fraction in state $|3\rangle$ for $140\,\mu$s RF pulses, except at
$750\,$G. Here, because of strong inelastic losses between
$|3\rangle$ and $|1\rangle$ atoms, we monitored the loss of atoms
in state $|2\rangle$ after applying RF sweeps of $3\,$ms duration
and $2\,$kHz width. All the data points are normalized to the same
atom number in state $|1\rangle$. The fit at low fields (solid
line) uses Eq.$\,$\ref{meanfield} with $n_1=2.2\times
10^{13}\,$cm$^{-3}$ and the theoretical calculations of the
scattering lengths. The error bars reflect uncertainty in the
state $|1\rangle$ atom number, and the uncertainty in the gaussian
fits to the spectra. The dashed line indicates the position of the
predicted $a_{13}$ resonance.} \label{fig:finaldata}
\end{figure}

\begin{figure}[htbf]
\begin{center}
\vskip0mm \epsfxsize=80mm {\epsfbox{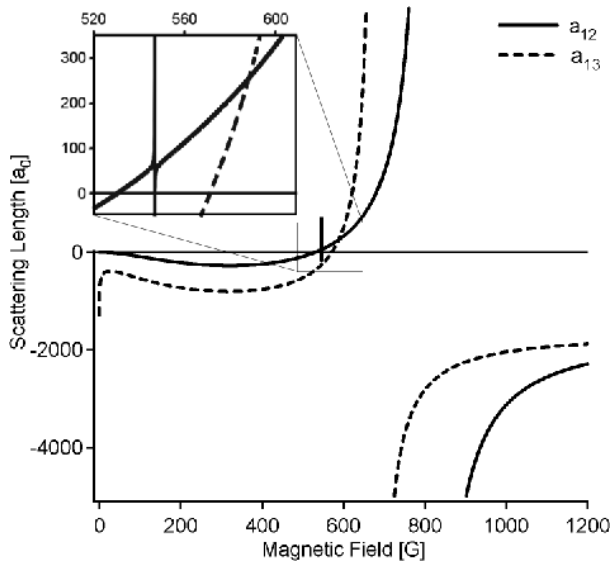}} \vskip1mm
\end{center}
\caption{s-wave scattering lengths $a_{12}$ and $a_{13}$ as a
function of magnetic field, obtained from a highly
model-independent quantum scattering calculation. The calculation
makes use of the presently available $^6$Li experimental data
[39] in a coupled channel approach to deduce
accumulated phases that characterize the less well-known short
range parts of the $^6$Li + $^6$Li scattering potential
[31]. $a_{12}$ has a narrow Feshbach resonance at
$550\, $G and a wide one at $810\, $G. $a_{13}$ has a wide Feshbach
resonance at $680\, $G. }
\label{fig:Feshbach}
\end{figure}

Fig.$\,$\ref{fig:finaldata} summarizes the results of our
mean-field measurements for a wide range of magnetic fields up to
$750\,$G. For magnetic fields up to $630\,$G, our data can be
explained fairly well using Eq.$\,$\ref{meanfield} with the
theoretical calculations of the scattering lengths shown in
Fig.$\,$\ref{fig:Feshbach}, and an effective density of $n_1=2.2
\times 10^{13}\,$cm$^{-3}$, consistent with the initial
preparation of the system at 500$\,$G. A narrow resonance of
$a_{12}$ at $\sim 550\,$G \cite{diec02,ohar02,vank03} is indicated
by the data, but was not fully resolved. We also see additional
structure near $470\,$G, which is not predicted by theory and
deserves further study.

The most important results of this paper are our observations for
fields above $630\,$G, where the measured shifts strongly deviate
from the predictions of Eq.$\,$\ref{meanfield}. In the region
between $630\,$G and $680\,$G, the two scattering lengths are
expected to be large and positive, with $a_{13} \gg a_{12}$
(Fig.\ref{fig:Feshbach}). Eq.$\,$\ref{meanfield} would thus
predict large positive mean-field shifts. In contrast, we observe
very small shifts, indicating almost perfect cancellation of the
two contributions. Even more surprisingly, we observe essentially
no mean-field shifts between $680\,$G and $750\,$G, where the two
scattering lengths are predicted to be very large in magnitude and
of opposite signs, and in a simple picture should add up to a huge
negative shift. These results are evidence for new phenomena in a
strongly interacting system, where the scattering length becomes
comparable to either the inverse wavevector of interacting
particles, or the interatomic separation.




Eq.$\,$\ref{meanfield} is valid only for low energies and weak
interactions, where the relative momentum of the two particles,
$\hbar k$, satisfies $k \ll 1/|a|$. For arbitrary values of $ka$,
the s-wave interaction between two atoms is described by replacing
the scattering length $a$ with the complex scattering amplitude:

\begin{equation}
\label{scat_amplitude} f= \frac{-a}{1+k^2a^2}(1-ika)
\end{equation}

The real part of $f$ determines energy shifts, and hence the
ground state properties of an interacting many-body system. The
imaginary part determines the (inverse) lifetime for elastic
scattering out of a momentum state, and hence the dynamic
properties of the system such as thermalization rates. For $k|a|
\rightarrow \infty$, the elastic cross-section $\sigma =
4\pi$Im$(f)/k$ monotonically approaches the well known ``unitarity
limited" value of $4\pi/k^2$. On the other hand, the two particle
contribution to the mean-field energy, proportional to $-$Re$(f) =
a/(1+k^2a^2)$, peaks at $|a|=1/k$, and then, counter-intuitively,
decreases as $1/|a|$ for increasing $|a|$. It has been shown
\cite{gehm03} that averaging Re$(f)$ over a zero-temperature Fermi
distribution with Fermi momentum $\hbar k_F$, limits its absolute
value to $1.05/k_{F}$, and drastically weakens its dependence on
the exact value of $a$ in the $k_{F}|a|>1$ regime. This results in
a prediction for the mean-field energy which is sensitive to the
sign of the scattering length, never exceeds $0.45 E_F$, and
remains finite for $k_F |a| \gg 1$. Hence, this approach could
qualitatively explain our results in the $630-680\,$G region, but
it is in clear contradiction with negligible resonance shifts in
the $680-750\,$G region \cite{scaling}.

We suggest that these discrepancies might be due to the fact that
we are in the high density regime, where $n|a|^3$ approaches
unity. In a degenerate Fermi gas, the interparticle spacing is
comparable to the inverse Fermi wavevector, $k_{F}^3\,=\,6 \pi^2
n$. Hence, the unitarity limit coincides with the breakdown of the
low-density approximation ($n|a|^3 \ll 1$) and higher-order
many-body effects can become important. Some recent many-body
calculations \cite{stee00,heis01,comb03} suggest that in the
regime $k_F |a| \gg 1$ (or $n|a|^3 \gg 1$), the interaction energy
is always negative and independent of both sign and magnitude of
$a$. This suggests that whenever the scattering length is large,
either positive or negative, the interaction energy is a universal
fraction of the Fermi energy \cite{gehm03}. This is a possible
explanation for the small line shifts that we observed for fields
above $630\,$G, where the interactions are strong in both states.

This picture is consistent with other recent experimental
observations \cite{rega03,gehm03,ohar02_2,bour03}. Expansion
energy measurements in a mixture of states $|1\rangle$ and
$|2\rangle$ of $^{6}$Li \cite{bour03}, showed a negative
interaction energy at 720$\,$G, which is on the repulsive side of
the predicted Feshbach resonance. RF spectroscopy in $^{40}$K
\cite{rega03} has also shown some saturation of the mean-field in
the vicinity of a Feshbach resonance, which may reflect the
unitarity limit.



In characterizing an interacting Fermi gas by RF spectroscopy, we
have demonstrated absence of clock shifts in a two-level system,
and introduced a three-level method for measuring mean-field
shifts. For strong interactions, we have found only small line
shifts which may reflect both the unitarity limit of binary
collisions and many-body effects. It would be very important to
distinguish between two-body and many-body effects by studying the
gas over a broad range of temperatures and densities. In a very
dilute and very cold gas, the weakly interacting limit could be
extended to very large values of $|a|$, thus allowing for direct
verification of molecular calculations. This presents experimental
challenges, because cooling changes the density and the
temperature together. It would also be interesting to study
similar phenomena in bosonic gases, in order to distinguish to
what extent the high density many-body effects depend on quantum
statistics. This new insight into the physics of strongly
interacting Fermi gases must be taken into account in the search
for superfluidity in these systems.




\begin{thebibliography}{10}
\bibitem{ODTload}
The transfer efficiency is limited by finite optical trap depth
and residual heating during the transfer.
\bibitem{density}
For our modest degeneracy, we can use a gaussian approximation for
the density distribution. The mean density is then lower than the
peak density by a factor of $2 {\sqrt 2}$.
\bibitem{clock23}
Note that due to fermionic antisymmetry, there is no clock shift
from interactions between atoms in states $|2\rangle$ and
$|3\rangle$. Mean field shifts only arise during a decoherence
time (see later in the text), but are negligible if the population
of atoms in state $|3\rangle$ is small.
\bibitem{scaling}
We note that for fields above $630\,$G, number losses generally
reduced $n_1$, and all the shifts in this region were scaled up by
a factor of $1.1-2.7$, but still remained negligible.
\end{thebibliography}
\end{document}